\documentclass[12pt]{article}

\newdimen\hsgraph \newdimen\vsgraph
\hsgraph=\hsize \advance\hsgraph by-1.2truein
\vsgraph=\vsize \advance\vsgraph by-1.5truein

\newcommand{\B}[1]{{\mathbb #1}}
\newcommand{\C}[1]{{\mathcal #1}}

\newcommand{\beq}{\begin{equation}}
\newcommand{\eeq}{\end{equation}}
\newcommand{\bea}{\begin{eqnarray}}
\newcommand{\eea}{\end{eqnarray}}
\newcommand{\rf}[1]{(\ref{#1})} 
\newcommand{\ket}[1]{\vert #1\rangle}
\newcommand{\bra}[1]{\langle #1\vert}

\newcommand{\ra}{\rightarrow}

\begin{document}
\topmargin 0pt
\oddsidemargin 5mm
\headheight 0pt
\topskip 0mm

\addtolength{\baselineskip}{0.20\baselineskip}

\pagestyle{empty}

\begin{flushright}
OUTP-99-61P\\
November 1999\\
hep-th/9911252
\end{flushright}

\begin{center}

\vspace{18pt}
{\Large \bf What does  $E_8$ know about $11$ dimensions ?}

\vspace{2 truecm}

{\sc Ian I. Kogan \footnote{e-mail: i.kogan@physics.ox.ac.uk} and 
  John F. Wheater\footnote{e-mail: j.wheater@physics.ox.ac.uk}}

\vspace{1 truecm}

{\em Department of Physics, University of Oxford \\
Theoretical Physics,\\
1 Keble Road,\\
 Oxford OX1 3NP, UK\\}

\vspace{3 truecm}

\end{center}

\noindent
{\bf Abstract.} We  discuss some possible relationships
in gauge theories, string theory and M theory in the light 
of some recent results obtained in gauge invariant supersymmetric
 quantum mechanics. In particular this reveals a new relationship
between the
 gauge group $E_8$ and 11-dimensional space.

\vfill
\begin{flushleft}
\end{flushleft}
\newpage
\setcounter{page}{1}
\pagestyle{plain}

A long time ago it was proposed by Eguchi and Kawai \cite{EK} that the dynamics of
a four dimensional $SU(N)$ lattice gauge theory at large $N$ could be described
by the same $SU(N)$ gauge theory on a single hyperube with periodic 
boundary conditions. Somehow the extensive nature of space time in such a
theory is not important, at least in the large $N$ limit.
  The Eguchi Kawai model turned out to have a number
of problems which required various modifications and it has never yielded 
a systematic way of solving lattice gauge theories in general. However the idea
is clearly appealing, especially in the context of modern developments
 in which  field theories  in different numbers of space-time dimensions are 
related by the compactification of some of the dimensions; the Eguchi Kawai
model is simply $SU(N)$ lattice gauge theory compactified on $T^4$ with the 
compactification radius sent to zero (or rather one lattice spacing). 
One practical  difficulty with this is precisely that in a lattice theory one
cannot continuously vary the compactification radius by changing the lattice
size; really we want to keep the lattice size fixed and vary the correlation
length so that the size of the box occupied by the continuum effective
theory is sent to zero. Perhaps it would be simpler to deal with a continuum
theory from the outset.  One might suspect that one crucial difference between
the modern compactification relations and the  Eguchi Kawai model is supersymmetry; 
this ensures that the energy of the ground state does not diverge
under compactification and seems (at least from the technical point of view)
to be a vital ingredient in demonstrating the network of dualities relating
different string theories and M-theory.

In a recent paper Kac and Smilga \cite{KS} have analyzed the zero mode structure of the 
supersymmetric quantum mechanics (SQM) obtained by dimensionally reducing
$D=3+1$ dimensional $\C N=4$ to supersymmetric Yang-Mills field theory
(SYM) with gauge group $G$ to $D=0+1$. In turn the $D=3+1$ theory can be regarded as the 
dimensional reduction of  $D=9+1$ $\C N=1$ SYM. The dynamics of this SQM had been first analysed
 in \cite{ch}, \cite{smilga}.
The SQM theory is known to have a continuous spectrum with states of all energies $E$ 
from zero (as guaranteed by supersymmetry) upwards; these states are not
normalizable. There is also a discrete spectrum of normalized $E=0$ states.
If the gauge group $G=SU(N)$ then it is known \cite{lightconesuper} 
 that the SQM Hamiltonian can be regarded as describing the regularized 
quantum 11-dimensional supermembrane (see for example \cite{townsend} and references therein)
 in the light-cone gauge just as   ordinary 
$SU(N)$ QM emerges in the  light-cone quantization of the bosonic membrane \cite{gh}. 
 It was shown in \cite{dln} that the spectrum of $SU(N)$ SQM is continuous
 which  kills the 
old  (i.e. first quantized ) supermembrane. However, this 
 is precisely the property
 which is necessary in a  matrix formulation of M-theory \cite{BFSS} where $N$ in $SU(N)$
 is   related to the number of `parton' $D0$ branes in a light-cone formulation
 of M-theory in a flat background.  More about M-theory and Matrix theory can be found in 
  \cite{witten}, \cite{M} and  \cite{matrix} and references therein.
 
If the SQM is to work as a formulation of M-theory then 
 it is crucial that there is only one $E=0$ ground state
and much effort has gone into investigating this. The proof was given
 for $SU(2)$ in
\cite{indexsu2}, and a lot of evidence accumulated for
 $SU(N)$  \cite{indexsun}.
 Kac and Smilga have now shown that this is indeed the case for all
 $SU(N)$ and  $U(N)$ gauge groups. 
However their results are much more far reaching and 
show that the other classical groups, $SO(N)$, $Sp(2r)$, and the exceptional groups, have 
a much richer structure of $E=0$ modes.

Consider a $D=d+1$ SYM theory with the $d$ spatial dimensions forming a compact manifold 
with isometry group $\C E$ and scale (compactification scale) size $\lambda$. 
We can dimensionally reduce the theory to SQM by sending $\lambda$ to zero. 
In doing so we lose the isometry group as a symmetry of the theory.
However we also create a vector space of $n_G$ $E=0$ modes which we will
denote $\{\vert 0,i\rangle,\, i=1\ldots n_G\}$; note that these
states are time-independent because $E=0$ and can therefore be regarded as
forming a real vector space not a complex one. A general
zero energy state can then be written
\beq 
\ket{ X}=\sum_{i=1}^{ n_G} X^i\ket{0,i}.
\label{one}
\eeq
The normalization condition applied to $\ket{ X}$ gives the constraint
\beq 
1=\sum_{i=1}^{ n_G} X^i X^i
\label{two}
\eeq
and thus the $X^i$ live on the sphere $S^{n_G-1}$. Thus the destruction of the original
isometry group  $\C E$ by the dimensional reduction is accompanied by the
emergence of a new one, $\C E' =SO(n_G)$. We propose that if $\C E\equiv\C E'$
then the process of compactification and reduction is continuous 
(ie the original $d$ dimensional manifold flows into the new one) and the
SQM is equivalent to the original field theory; in this sense such models
provide a realization of the Eguchi Kawai idea. 

>From the rules given by Kac and Smilga we list in Table 1 the simple groups
which have (possibly) physically interesting values of $n_G$.
\begin{table}
\begin{center}
\begin{tabular}[t]{|c|l|}
\hline
$n_G$& Gauge group  $G$\\ \hline
2&$SO(8)$, $SO(9)$, $SO(10)$, $SO(11)$, $Sp(6)$, $Sp(8)$, $G_2$\\ \hline
3&$SO(12)$, $SO(13)$, $SO(14)$,  $Sp(10)$, $E_6$\\ \hline
4&$SO(15)$, $Sp(12)$, $F_4$\\ \hline
5&$SO(16)$, $SO(17)$, $SO(18)$,  $Sp(14)$\\ \hline
9&$SO(23)$\\ \hline
10&$Sp(20)$\\ \hline
11&$SO(24)$, $E_8$\\ \hline
\end{tabular}
\end{center}
\end{table}
If we dimensionally reduce $D=3+1$ $\C N=4$ SYM compactified on $S^3$
then we see from the table
that there are three possible simple $G$ with $n_G=4$ which will reproduce
the $S^3$.  In addition there is a substantial number of direct products
made up from pairs of groups with $n_G=2$. 
For $D=9+1$ $\C N=1$ SYM compactified on $S^9$ there 
is only one possible simple group, $Sp(20)$, but again many
direct product groups with one member taken from the $n_G=2$ list and the other
from the $n_G=5$ list.

The groups involved in this dimensional reduction do not seem to
 have any particular  physical significance; it could be pure numerology, just
a mathematical game. However,
let us now take $ D = 10 +1 = 11$ and we see there are two groups $SO(24)$ and
 $E_8$. We  have here
 a very surprising and striking  fact  - the group $E_8$ {\it knows} about
 $11$ dimensions!

There is  a natural place for $E_8$ in string theory - through the  heterotic 
construction and Narain compactification \cite{heterotic} -
 roughly speaking this appears
 as a group  with self-dual and even root lattice
\footnote{We are not going to discuss $SO(24)$ beyond pointing out that
it
is the spatial symmetry group of the critical bosonic  string in
 light-cone gauge but the relation  between  11 dimensions  and  bosonic strings is 
unclear to us  for now.}. The fact that these 
lattices exist in dimensions  $0(mod8)$  fits 
nicely with the fact that the 
critical dimensions of bosonic and fermionic strings are 
26 and 10 and the  difference fits the group $E_8 \times E_8$ (also $SO(32)$ of course, but for $SO(32)$
 one may have open strings  contrary to $E_8 \times E_8$). The fact 
that this group can be obtained
  in closed string theory which is finite guarantees that the
 Green-Schwarz mechanism \cite{GS}
of anomaly cancellation works. Let us note that in Narain toroidal compactification of heterotic
 string on a lattice $\Gamma_{26-d, 10-d}$ with $d<10$ \cite{narain} one may  get other groups -
  but in this case the space has a smaller symmetry $SO(d)$.

Now consider the  Horava-Witten construction \cite{HW} of heterotic M-theory.
  Here  the ten-dimensional $E_8\times E_8$ 
heterotic string is related to an eleven-dimensional theory on the orbifold 
$R^{10}\times S^1/Z_2$ and the presence of 10-dimensional boundaries of 11-dimensional
 space leads to the existence of an $E_8$ gauge group on each boundary in order
to cancel diffeomorphism anomalies.
However  it  seems that $E_8$  is not directly related to 11 dimensions and  knows nothing about
 the  maximal  Lorentz group $SO(11)$. 
But now we should take account of this new information about  
 $E_8$ SQM.
  Consider heterotic M-theory on a space
 $R^{1}\times T^9\times S^1/Z_2$ - each $E_8$ gauge theory is on a 9-torus $T^9$ and the $R^1$ factor is time. Let
the radii of the torus be  $R_i = \lambda_i R, i=1\ldots 9$ where 
the $\lambda_i$ are conformal factors
and consider the limit $R \ra 0$.
 We get a $1+1$ dimensional theory on the orbifold $S^1/Z_2$. 
This theory contains the conformal factors of  $T^9$ 
whose quantum mechanics is 
trivial, and two-dimensional gravity with a copy of  $E_8$ SQM at
each of the two 
singular points on the orbifold. It is tempting to argue that the
two-dimensional gravity
 is non-dynamical  and therefore we are just left with the two copies
of $E_8$ SQM each with an $SO(11)$ isometry group 
on their zero energy normalizable subspace; note that this would 
not depend on
the value $R_{11}$. If $R_{11}$ is small we have weakly coupled
 string theory compactified on a shrinking torus,
 and if it is large we have  the  strong coupling limit.
 However it seems to us that the conclusion that there
are two $SO(11)$ isometry groups  cannot be true 
and that in fact the gravity must somehow mediate a coupling between the 
boundaries so that the quantum theory has only one $SO(11)$ symmetry.
Our reason for this is a string theory argument that we give in the 
next paragraph.


 String theory on $T^9$ which
 is shrunk to a point is  $T$-dual to an infinitely large torus, i.e. essentially $R^9$. Since the theory on $T^9$ is heterotic so is the $T$-dual theory
on $R^9$ (in contrast 
 to  non-heterotic M-theory where $d=11$ supergravity corresponds to
 the  strongly coupled $IIA$ string \cite{witten} and the T-dual  transforms 
   it into type $IIB$). 
    Now  under $T$-duality we have that,
 \beq
 R \ra \tilde {R}= {1\over R}, ~~~ g_s \ra \tilde{g}_s = g_s \sqrt{\frac{\tilde {R}}{R}} =
 \frac{g_s}{R}
\eeq 
so when $R \ra 0$ we see that $\tilde{R} \ra \infty$ and $\tilde{R}_{11} 
\sim \tilde g_{s} \ra \infty$. Thus  string theory arguments
 show that we should  recover the  $SO(11)$ symmetry.
 However if we started from strong coupling from the
 very beginning we can not apply string 
arguments, but this is  precisely the situation discussed in the
previous paragraph.

The group $SO(11)$ we get is hypothetically the full Lorentz group of M-theory.
 Because the membrane Lorentz algebra is defined in  light-cone gauge we have to check that
 there is  full Lorentz invariance just as in a light-cone string theory. The
classical
 Lorentz  algebra  becomes closed only in the  $N \ra \infty$ limit, and
quantum-mechanically it is still  unknown if $D=11$ is a critical dimension.
 Although the full quantum commutator is still unknown,  it has been
shown that the lowest  non-trivial anomalous terms in the 
 commutators $[M^{-~i}, M^{-~j}]$
are zero \cite{so11}. The
 $SO(11)$ should be an {\it exact} quantum symmetry. In the case 
of $R^{10}\times S^1/Z_2$ there is an obvious $SO(10)\times P(10)$ 
where $P(10)$ denotes the
Poincar\'e symmetry. The full $P(11)$ is broken by the existence of 
the orbifold planes.  This group is nothing but a contraction of the
full $SO(11)$; in a picture of two concentric 10-spheres the full 
$SO(11)$ acts faithfully on the space between them.

If we  consider matrix theory on a  nine-dimensional thorus we have instead
of $0+1$ SQM a fully fledged
 $9+1$ SYM - again this theory only has a maximal $SO(10)$ Lorentz symmetry, but if we add
 $9+1$ dimensional $E_8$ theory and reduced the system as a whole again to $0+1$ the full $SO(11)$ symmetry
  will be produced by the  $E_8$ SQM.
   
We have argued that the {\it two} $E_8$s should produce
only  one  $SO(11)$. Suppose this is wrong; is there any other 
interpretation? A generic state for $E_8 \times E_8$ theory
 has the form
\beq 
\ket{X,Y}=\sum_{i,j=1}^{11} X^i Y^j\ket{0,i}_1 \ket{0,j}_2 = \ket{X}_1\ket{Y}_2
\label{twoe8}
\eeq
which looks like we have $R^{11} \times R^{11} = R^{121}$ - not at all
what we are looking for.
 A possible loophole is to consider instead of states $\ket{X}_1\ket{Y}_2$
 operators $\ket{X}\bra{Y}$. This simply means that the second Hilbert space we consider
 as a conjugate to the first one and the quantum mechanical description is given by a {\it
 density matrix}
\beq
\rho (X,Y) = \sum_{i,j=1}^{11} X^i Y^j\ket{0,i} \bra{0,j}
\eeq
 This is actually a very nice picture because so far we have no idea  how even to start to
  construct dynamics on this 10-dimensional sphere. But now if we have the space of non-trivial ground states in $E_8$ providing coordinates, 
  and momenta being the cotangent bundle to this space, 
 we can at least formulate the canonical symplectic form
   \beq
    \Omega = dp^i \wedge dq^i
    \eeq
    and start to develop the dynamics. For normalised wave functions  the phase space is going to be
     $S^{10} \times T^* S^{10}$,  so two
 boundaries of 11-dimensional space in heterotic M-theory
      play the role of two coordinates in the density matrix $\rho(x+\eta,x-\eta)$ a phase space of some dual theory which has exact $SO(11)$ symmetry.
The difference between the coordinates $\eta$ is parametrised by the tangent
bundle to $S^{10}$ and one can see from a Wigner function that a momentum
is dual to this difference which justifies the structure of phase space.
     How to find a Hamiltonian on this phase space and formulate a dynamics 
is a  question which remains to be 
     answered!

We would like to thank Andr\'e Smilga for telling us about the results
of \cite{KS} and the  PPARC Fast Travel fund for supporting his visit here.

\end{document}